\documentclass[12pt]{article}
\usepackage{amsmath,amssymb,graphicx} 
\def\be{\begin{equation}}
\def\ee{\end{equation}}
\def\bea{\begin{eqnarray}}
\def\eea{\end{eqnarray}}
\def\ba{\begin{array}}
\def\ea{\end{array}}
\def\bal{\begin{align}}
\def\eali{\end{align}}
\def\bsu{\begin{subequations}}
\def\esu{\end{subequations}}
\def\bib{\bibitem}
\def\ql{\lq\lq}
\def\qr{\rq\rq}
\def\fn{\footnote}
\def\ef{\eqref}

\def\lp{\left(}
\def\rp{\right)}
\def\nn{\nonumber}

\def\bq{\begin{quote}}
\def\eq{\end{quote}}

\def\pt{\partial}
\def\ind{\hspace{.5cm}}
\def\hl{\rule[-0.1cm]{7cm}{0.02cm}\\}
\def\nc{\newcommand}
\nc{\prt}[2]{\frac{\partial{#1}}{\partial{#2}}}
\def\half{{\tfrac{1}{2}}}

\def\a{\alpha}
\def\b{\beta}

\def\G{\Gamma}
\def\d{\delta}

\def\k{\kappa}
\def\l{\lambda}

\def\m{\mu}
\def\n{\nu}
\def\p{\pi}
\def\r{\rho}
\def\t{\tau}

\nc{\chrd}[3]{(g_{{#1}{#3},{#2}}+g_{{#3}{#2},{#1}}-g_{{#1}{#2},{#3}})}
\nc{\chru}[3]{\{\!\!\begin{array}{c}{{}_#1}\\{{}^#2}{{}^#3}\end{array}\!\!\}}
\nc{\gamu}[3]{{\G^{#1}}_{{#2}{#3}}}
\nc{\chrud}[3]{\left\{\!\!\begin{array}{c}{#1}\;\;{#2}\\{#3}\end{array}\!\!\right\}}
\begin{document}
\begin{center}
\Large{\bf How Einstein Discovered Dark Energy}\\
\vspace{.5cm}
\large{Alex Harvey}$^{a)}$ \\
\vspace{.15cm}
\normalsize
Visiting Scholar \\
New York University \\
New York, NY 10003 \\
\vspace{.8cm}
\end{center} 
\vspace{.8cm}
\begin{abstract}          
In 1917 Einstein published his {\it Cosmological Considerations Concerning the General Theory of Relativity.}  In it was the first use of the cosmological constant.  Shortly thereafter Schr\"{o}dinger presented a note providing a solution to these same equations with the cosmological constant term transposed to the right hand side thus making it part of the stress-energy tensor.  Einstein commented that if Schr\"{o}dinger had something more than a mere mathematical convenience in mind he should describe its properties.  Then Einstein detailed what some of these properties might be.  In so doing, he gave the first description of Dark Energy.  We present a translation of Schr\"odinger's paper and Einstein's response.
\end{abstract}
\section{Introduction}
In 1917 Einstein published his {\it Cosmological Considerations Concerning the General Theory of Relativity\/} \cite{ein1}.  The field equations he used included the cosmological constant.
\be\label{a}     
              R_{\m\n} - \l g_{\m\n} = - \k(T_{\m\n} -\half g_{\m\n}T) \,.
\ee
Shortly thereafter Schr\"{o}dinger presented a note \cite{schro} in which he transposed the cosmological constant term to the right hand side thus making it part of the source, that is, part of the stress-energy tensor.  
\be\label{b}
                  R_{\m\n} = \l g_{\m\n} - \k(T_{\m\n} -\half g_{\m\n}T)
\ee   
Einstein immediately recognized the implications and responded with an analytic comment \cite{ein2}.  If the $\l$ term were to remain a constant then the shift was trivial.  If, however, it was truly intended to be a source then its properties would necessarily have to be specified.   There is no record of a response by Schr\"{o}dinger to Einstein's comment and there the matter remained.  The exchange between Einstein and Schr\"{o}dinger, less technical details, is described by Mehra and Reichenberg \cite{mehra}.

A translation of Schr\"{o}dinger's paper follows directly.  In it all equations, footnotes, and abbreviations are as in the original. Explanatory comments are interpolated in the text.  Footnotes and citations are collected at the end and where obscure, they are expanded. 
\section{Schr\"{o}dinger's Paper}
\vspace{0.5cm}
\begin{center}
\large{Concerning a System of Solutions of the\\Generally Covariant Equations for Gravitation}
\end{center}
\setcounter{equation}{0}
Same time ago Herr Einstein${}^1$ presented a system of energy components for matter, $T^{\m\n}$, and gravitational potentials, $g_{\m\n}$, which integrate, precisely, the generally covariant field equations. He suggested that this system approximated the real structure of space and matter in the large. Briefly stated, it describes incoherent matter of constant density at rest that fills${}^2$ a closed spatial continuum of finite volume having the metric properties of a hypersphere.

For the field equations, a slightly different form than the original one is assumed.${}^3$

\hl
\small{The \ql slightly different form\qr is simply the transposition of $\l g_{\m\n}$ from the l.h.s to the r.h.s. thereby obtaining Eq. \ef{b} (above).}\normalsize \vspace{-4mm}\\ 
\hl

The concept of a closed total space seems to me to be of extraordinary importance for general relativity. This is not only, and not mainly, because of the \ql desolation argument\qr from which Herr Einstein started. I see its principle value rather in the fact that by development of this thought general relativity promises to be what its name says and what it — according to my view — only formerly has been, so to say on paper.${}^4$

\hl \small{It is not clear what Schr\"{o}dinger meant by "desolation argument". It is possible  that he is referring to the problem of constructing a cosmological model within the context of Newtonian gravitation.}\normalsize \vspace{-4mm}\\ 
\hl

Under these conditions it is also not without interest to remark that the completely analogous system of solutions exists in its original form — without the terms added by Herr Einstein {\it l.c.}. The difference is superficial and slight: The potentials remain unchanged, only the energy tensor of matter takes a different form. 
\hl
 \small{This refers to the transfer of $\l g_{\m\n}$ to the right hand side.\normalsize \vspace{-4mm}\\ 
\hl

The system of $g_{\m\n}$ is also [compare Einstein l.c. equations (7), (8) and (12)]:

\bea
                       g_{44} &=& 1,\;\; g_{14} =g_{24}=g_{34} = 0  \\
g_{\m\n} &=& -\lp \d_{\m\n} + \frac{x_\m\,x_\n}{R^2 -x_1^2 -x_2^2 -x_3^2}\rp  \nn\\
\m,\n &=& 1,\,2,\,3,\;\; \d_{\m\n} = \begin{array}{c}0\\1\end{array} \; {\rm if}\: \n \left\{\begin{array}{c}
\neq \\ = \end{array}\right\} \m \nn \,.
\eea

If one now assumes a mixed energy tensor of matter with vanishing off-diagonal components${}^5$
\be
T^\m_\n =\lp \begin{array}{cccc} T^1_1 &0&0& 0\\0&T^2_2 &0&0\\0&0&T^3_3&0\\0&0&0&T^4_4\end{array}\rp
\ee
the field equations becomet
\bea
-\frac{\pt}{\pt x_\a}\chrud{\m}{\n}{\a} +\chrud{\m}{\a}{\b}\chrud{\n}{\b}{\a}\nn\\
   -\frac{\pt^2\log\sqrt{-g}}{\pt_\m\pt_\n} 
        - \chrud{\m}{\n}{\a}\frac{\pt\log\sqrt{-g}}{\pt x_\a} &=&-\k(T_{\m\n} -\half\, g_{\m\n}T)
\eea
and are satisfied only if
\be
          T^1_1\equiv  T^2_2\equiv  T^3_3\equiv  T^4_4\equiv \frac{1}{\k R^2} \equiv const 
\ee

\flushleft\hl\small{The l.h.s. is just $R_{\m\n}$.  The Christoffel symbols were written this way at that time.}\normalsize \vspace{-4Mm}\\ 
\hl 
\vspace{3mm}

The calculation does not present any difficulty; it proceeds exactly as with Einstein {\it l.c.\/} Most importantly, from a  logical point of view, one might remark: $k$ and $R$ must naturally be specified to be universal constants to start with, i.e., before one writes potentials (1) and the field equations (3) respectively.  

\hspace{.5cm}Because the calculation can be carried out easily only for points with the property $x_1=x_2=x_3=0$, condition (4) is satisfied only for those points identical in $x_4$!  For the field equations to be satisfied identically it is necessary, permissible and,  sufficient to require, also, the identical validity of (4) for the spatial coordinates.  In this way these relations become invariant with respect to arbitrary transformations of the spatial coordinates $x_1,\,x_2,\,x_3$ among themselves and the time $x_4$ into itself. Through simple transformations of this kind one can move any arbitrary point into one of the two \ql$x_4,\,axes$\qr.

\hspace{.5cm}It is desirable to connect the system of solutions with a reasonable physical example.  This is possible to a certain degree if one accepts Einstein's hypothesis for the energy tensor of a continuous compressible fluid.  In this case we have${}^6$
\bea
                    T^\n_\m &=&-\d^\n_\m\, p + g_{\m\a}\frac{dx^\a}{ds}\frac{dx^\n}{ds}\,\r \nn \\
            \d_\m^\n &=& \left . \begin{array}{c}0\\1\end{array}\right\} \; {\rm if}\: \n \left\{\begin{array}{c}
\neq \\ = \end{array}\right\} \nn\m \,.
\eea
where $p$ are $\r$ scalars, the \ql pressure\qr and density of the fluid. If one sets
\be
\frac{dx_1}{ds} \equiv \frac{dx_2}{ds} \equiv \frac{dx_3}{ds} \equiv 0,\;\frac{dx_4}{ds} \equiv 1 
\ee
which is, because of (1), the basic equation
\be
                          ds^2 = g_{\m\n}dx^\m dx^\n 
\ee
and with the equations of motion (that is, a geodesic)
\be
    \frac{d^2 x_\t}{ds^2} + \chrud{\m}{\n}{\t}\frac{dx_\m}{ds}\frac{dx_\n}{ds} = 0 
\ee
we obtain
\be
            T^\n_\m = \lp \begin{array}{cccc}
                            -p & 0 & 0 & 0 \\
                            0 & -p & 0 & 0 \\
                            0 & 0 & -p & 0 \\
                            0 & 0 & 0 & p -\r  \end{array} \rp 
\ee

\hl\small{In transposing $\l$ to the r.h.s., $\l$ is re-identified as a negative pressure.}\normalsize\vspace{-4mm}\\ 
\hl
${}$
                              
This scheme differs (with suitable values for $\r$ and $p$) only superficially from our 
{\it hypotheses\/} above [(2) and (4)] for the energy tensor.

These hypotheses see matter in the large essentially as a compressible fluid of constant density at rest under a constant spatially isotropic tension which according to (4) must be equal to $1/3$ of the rest density of energy.${}^7$

From the point of view of the old theory, which for Einstein plays the role of a first approximation,\flushleft\hl\small{ 
What is meant here is not clear.  Possibly it refers to the field equations without the 
$\l$ term.}\normalsize \vspace{-4mm}\\ \hl
${}$\\                         
It is not at all disturbing that a tension of this sign exists in a matter thought to be present and uniformly spread over immense spaces and acting on itself under Newton's law of force.  On the other hand, on closer scrutiny, the following fact is astonishing: Satisfaction of the 10th field equation (equation [3], $\m = \n = 4$) is due to the fact that the expression
\be
                      T^4_4 - T^1_1 - T^2_2 - T^3_3 
\ee 
vanishes. It now plays, in first approximation, the role of the Newtonian gravitating mass density, for which as is well known, only the 1O${}^{th}$ field equation needs to be considered.

\flushleft\hl\small{To obtain the Newtonian limit of the field equations, only the 4-4 equation is required.}\normalsize \vspace{-4mm}\\ \hl 
${}$\\
  The vanishing of the \ql mass-term\qr (10) seemed to me, initially, to be questionable as to the utility of the given system of solutions as a representation of matter in the large.  But now I find just this result rather satisfactory Firstly, one has to expect, if not demand, from a theory in which the motion of mass is relative, i.e. determined only through the interaction of masses, that in any case the total mass of the universe vanishes.  

That the mass-term in our case vanishes not only {\it in toto\/} but at every discrete point is a consequence of the fiction that matter is exactly at rest and completely, uniformly spatially distributed. It seems to me completely in the spirit of the relativity of mass to assume that the interaction function that we call inertial or gravitational mass appears or comes about only by deviations from the uniform static distribution. No harm is done when the uniform static distribution  in which the mass can neither act inertial nor gravitating, yields a vanishing mass-term.

Naturally, there is now the task to obtain the real empirical processes for individual concrete cases, so to say, by variation of this very uniform \ql inert\qr integral system. Then - and only then — I feel general relativity would fulfill those postulates that were for its origin the logical 
reasons.${}^7$  By the way, a reminder, this is also the value of the light pressure; but the sign is opposite! 

According to equation (4) the quantity $1/\k$ has the following physical meanings
\bea
     a)\; \frac{1}{\k} &=& \frac{1}{6\p^2}\frac{2\p^2 R^3T^4_4}{R} = \frac{1}{6\p^2}\frac{E}{R} \nn\\
     b)\; \frac{1}{\k} &=& \frac{1}{4\p}\cdot 4\p R^2 T^1_1 =\frac{1}{4\p}p  \nn
\eea
where $E$ is the total energy of the cosmos, $R$ (as above) the radius, $p$ the one-sided tension in an equatorial plane.  For the last results (with $\k = \frac{8\p k^2}{c^2}g^{-1}cm$, $k^2=6.68\cdot 10^{-8}gm^{-1}cm^3 sec^{-2}$; $c=3\cdot 10^{10}cm\;sec{-1}$;
\bea
  p&=&\frac{4\p}{\k} = \frac{c^2}{2\k^2}g\;cm^{-1}\;\; {\rm that\;\; means}\;\; g\; {\rm cm\;light\; sec}^{-2} \nn\\
        &=&\frac{c^4}{2\k^2} g\;  cm \; sec^{-2} = 6.06\cdot 10^{18} dyne \,. \nn
\eea
\hl
${}^1$A. Einstein, \ql Kosmologische Betrachttungen zur allgemeinen Relativit\"atstheorie\qr, {\it Sitzungsberichte der Preussischen Akademie der Wissenschaften\/} 142-150 (1917), Berlin. In what follows it will be cited by l.c.   \\ \vspace{2mm}
${}^2$Perhaps it is correct to say: \ql constructed\qr, (instead of fulfilled) \\ \vspace{2mm}
${}^3$A. Einstein, \ql Die Grundlage der Allgemeine 
Relativit\"{a}tstheorie\qr, {\it Annalen der Physik\/}, {\bf 49}, 769 (1916).  An English translation appears in {\it The Collected Papers of Albert Einstein\/}, vol. 6, A. Engel, translator,          E.L.Schucking, consultant, 146-200, (1999) Princeton University Press, Princeton, NJ.\\  \vspace{2mm}
${}^4$Compare especially to Einstein, l.c., p. 147 [The ds Sitter citation is obscure. It may refer to Koninklike Akademie van Wettenschappen, 1217-1224 (1917) which contains comments on Einstein's Kosmologische Betrachtungen."  \\ \vspace{2mm}
${}^5$If one does not make this assumption, it will appear as a requirement after Eq. (4). \\
${}^6$See  A. Einstein, Die Grundlage ... p. 52. [This reference is unclear.  The pagination of Einstein's article is 769-822.] \\ \vspace{2mm}
${}^7$ This is just the negative of the pressure of light.\\ \vspace{2mm}
${}^8$I thank Herr Flamm for many discussions to clarify these concepts.  \\ \vspace{2mm}
${}^9$Compare this with A. Einstein, Die Grundlage ... , Sect. (2).
\section{Einstein's Response}
\vspace{0.5cm}
\begin{center}
\large{Comment on Schr\"{o}dinger's Note \ql Concerning a System of Solutions of the Generally Covariant Equations for Gravitation\qr}
\end{center}
Einstein begins with
\bq
\ql When I wrote my description of the cosmic gravitational field I naturally noticed, as the obvious
possibility, the variant Herr Schr\"{o}dinger had discussed. But I must confess that I did not consider this interpretation worthy of mention.\qr
\eq
Later, he finishes with
\bq 
\ql The author [Schr\"{o}dinger] is silent about the law according to which $p$ should be determined as a function of the coordinates. We will consider only two possibilities: \\
\vspace{3mm}
1. $p$ is a universal constant. In this case Herr Schr\"{o}dinger's model completely agrees with mine. In order to see this, one merely needs to exchange the letter p with the letter $\l$ and bring the corresponding term over to the left-hand side of the field equations. Therefore, this is not the case the author could have had in mind.\\  
\vspace{3mm}
2. $p$ is a variable. Then a differential equation is required which determines $p$ as a function
of x1 . . . x4. This means, one not only has to start out from the hypothesis of the existence of a nonobservable negative density in interstellar spaces but also has to postulate a hypothetical law about the space-time distribution of this mass density.  
\ind The course taken by Herr Schr\"{o}dinger does not appear possible to me because it leads too deeply into the thicket of hypotheses.\qr
\eq

Remarkably, in  the second possibility, Einstein described not only the central problem of the search for dark energy but the headaches in formulating its structure.  Of course, this occurred long before the advent of quantum-field theoretic concerns about the 120 orders of magnitude disparity between the observed vale of the cosmological constant and the value of the zero-point energy as well as the later discovery of the Sn(1a) supernovae with their implications, so the discussion vanished into the archives.  Although the exchange disappeared from sight, the cosmological constant did not.  It was used by many cosmologists for various purposes.  (See Harvey \cite{harvey} for examples, discussion, and an extensive bibliography.)
\section{Dark Energy}
In the late 1990's, Einstein's concern about a \ql thicket of hypotheses\qr became real.  Observations on Sn(1a) supernovae by Reiss {\it et al\/} \cite{riess} and Perlmutter {\it et al\/} \cite{perl} indicated that the expansion of the universe was accelerating.  This acceleration may be explained in either of the 2 ways that parallel Einstein's comments to Schr\"{o}dinger: the cosmological constant is either a differential-geometric term in the field equations or a manifestation of a \ql field\qr driving the acceleration.  In the former, it is simply an observable constant; in the latter it must be extracted from Einstein's \ql ... thicket of hypothesis\qr.  This field is called \ql quintessence\qr.  Quintessence is a dynamic scalar field with potential energy tuned to provided appropriate dark energy at different  epochs of cosmic evolution.  
\fn{See Zlatev \cite{zlatev} or Linder \cite{linder} for discussions of specific models. For a broad discussion see Carroll \cite{carroll}.  For a history of the cosmological constant see See Straumann \cite {straumann}.}   The choice between regarding the cosmological constant as either a term in the field equations or a manifestation of dark energy is exacerbated by the circumstance that evidence for the existence of the latter has yet to be found.
\section{Epilogue}
Although Schr\"{o}dinger never again discussed the cosmological constant Einstein did.  Going from the cosmological to the infinitesimal he utilized it in an attempt to cope with what was at the time a very important problem.  Recall that in 1918 the only elementary particles known were the electron and the proton.  Physicists were attempting to understand why these were stable 
despite their internal electromagnetic repulsion.  Most attempts were based solely on electromagnetic theory.  For a review of these efforts see Pauli \cite{pauli}.  Einstein's effort was to construct a model in which stability was achieved through the use of gravitational forces.  In particular, he used modified gravitational field equations which included the cosmological constant \cite{ein4}.  The attempt was not successful and this was the last time he mentioned the cosmological constant other than to denounce it.
%
\section{Acknowledgment}
My deep thanks to Prof. Engelbert Schucking for suggesting this project as well as providing a translation of Schr\"{o}dinger's paper.\\
\vspace{4mm}
${}^{a)}$Prof. Emeritus, Queens College, City University of New York; ah30@nyu.edu.\\
\vspace{4mm}
%
  
\end{document}